\documentclass[]{interact}
\usepackage{epstopdf}% To incorporate .eps illustrations using PDFLaTeX, etc.
\usepackage[caption=false]{subfig}% Support for small, `sub' figures and tables
\usepackage{xcolor}
\usepackage{float}  
\usepackage{lmodern}  
\usepackage{multicol}
\usepackage[numbers,sort&compress,merge]{natbib}% Citation support using natbib.sty
\bibpunct[, ]{[}{]}{,}{n}{,}{,}% Citation support using natbib.sty
\renewcommand\bibfont{\fontsize{10}{12}\selectfont}% Bibliography support using natbib.sty

\theoremstyle{plain}% Theorem-like structures provided by amsthm.sty

\theoremstyle{definition}

\theoremstyle{remark}

\begin{document}

%\articletype{ARTICLE TEMPLATE}% Specify the article type or omit as appropriate

\title{Isomer- and state-dependent ion--molecule reactions between Coulomb-crystallised Ca$^+$ ions and 1,2-dichloroethene$^\dag$}

\author{
\name{Lei Xu\textsuperscript{a}, Richard Karl\textsuperscript{a}, Jutta Toscano\textsuperscript{a}, and Stefan Willitsch\textsuperscript{a}
\thanks{\dag Dedicated to Prof. Fr\'ed\'eric Merkt on the occasion of his 60$^\text{th}$ birthday.}
}
\thanks{Corresponding author: Stefan Willitsch, stefan.willitsch@unibas.ch}
\affil{\textsuperscript{a}Department of Chemistry, University of Basel, Klingelbergstrasse 80, 4056, Basel, Switzerland}
}

\maketitle

\begin{abstract}
We report a systematic investigation of isomer- and state-dependent reactions between Coulomb-crystallised laser-cooled Ca$^+$ ions and \emph{cis/trans}-1,2-dichloroethene (DCE) isomers. By manipulating the electronic state populations of Ca$^+$ through tuning of laser cooling parameters, we observed distinct reactivities in its ground and excited states, as well as with the geometric isomers of DCE. Our experiments revealed two primary reaction channels, formation of CaCl$^+$ and C$_2$H$_2$CaCl$^+$, followed by secondary reaction pathways. While excited-state reactions proceed at rate coefficients consistent with capture theory predictions, ground-state reactions show a systematically lower reactivity. \emph{Ab initio} calculations of reaction pathways suggest that this suppression stems from the formation of long-lived reaction complexes. The \textit{cis} isomer was found to exhibit a higher reactivity with all electronic states of Ca$^+$ than its \textit{trans} counterpart. The present study provides insights into the combined effects of molecular structure and quantum states influencing ion--molecule reaction dynamics.

% \vspace{1em}
% \begin{center}
% \includegraphics[width=0.8\textwidth]{DCE+Ca+_abstract_ Fig.pdf}
% \end{center}
% \vspace{1em}
\end{abstract}


%%%MAIN TEXT%%%%

\section{Introduction}

Gas-phase ion--molecule reactions represent fundamental processes in chemistry which play important roles in many chemical environments ranging from interstellar media~\cite{mcguire20a} to planetary atmospheres~\cite{shuman15a} and plasmas~\cite{williams02a}. Studying these reactions on the quantum level helps to unravel fundamental aspects of chemical reactivity and dynamics~\cite{meyer17a}. Established experimental techniques, such as selected ion-flow tubes (SIFT)~\cite{adams76a}, guided ion beams~\cite{armentrout02a} and supersonic flows (CRESU)~\cite{potapov17a} have yielded valuable insights into reaction kinetics and mechanisms, though each of these approaches presents inherent limitations for state-selective studies.

Over the past years, new experimental techniques have emerged which enable the study of ion--molecule processes under highly controlled conditions. Merkt and co-workers developed an innovative merged-beam technique with which ionic processes can be studied in the orbit of a Rydberg species at very low collision energies~\cite{allmendinger16a, allmendinger16b} while quantum states can be manipulated using microwaves~\cite{martins25a}. Additionally, Coulomb crystals of cold ions~\cite{willitsch12a, willitsch17a} such as Be$^+$~\cite{roth06b,yang18a}, Mg$^+$~\cite{molhave00a,staanum08a}, Ca$^+$~\cite{willitsch08a, gingell10a,chang13a,schmid19a,kilaj23a}, and Ba$^+$~\cite{roth08a,hall13b,mahdian21a,mohammadi21a} in radiofrequency (RF) ion traps offer an attractive platform for probing quantum-state-specific reaction dynamics. In these systems, the ions are laser-cooled to millikelvin temperatures and arrange in ordered structures in the trap thus enabling precise quantum state manipulation \textit{via} laser excitation~\cite{willitsch17a}. Ion--molecule reactions are often barrierless and thus described using long-range capture models which are, in principle, independent of the state of the ionic reactant~\cite{zhelyazkova20a,tsikritea22a,clary84a}. However, previous studies have often revealed significant differences in reactivity between the ground and excited electronic states~\cite{roth06b,yang18a,molhave00a,staanum08a,gingell10a,chang13a,schmid19a,kilaj23a,roth08a,hall13b,doerfler19a, Kocheril2025a}. 

Beyond the influence of specific quantum states, molecular structure plays an important role in determining reaction dynamics. For instance, conformational and geometric, i.e., \emph{cis-/trans-}, isomers can exhibit significantly different reactivities despite their identical chemical compositions~\cite{chang13a, schmid20a, kilaj21a,ploenes24a,richardson24a,zagorecmarks24a}. In this context, 1,2-dichloroethene (C$_2$H$_2$Cl$_2$, DCE) represents an interesting model system for investigating isomer-specific effects. The \emph{cis-} and \emph{trans-} isomers of DCE exhibit markedly different molecular properties: while the \textit{trans} species is apolar, the \textit{cis} isomer possesses a dipole moment of 1.9~D~\cite{haynes16a}, leading to different long-range interactions with ions. The reactivity of the isomers of DCE has been the subject of several previous investigations including charge-transfer~\cite{zhi21a}, SIFT~\cite{mikhailov06a,rebrion88b} and CRESU experiments~\cite{rebrion88b}, which render it a well-characterised benchmark system.

In this study, we investigate reactions between Coulomb-crystallised Ca$^+$ ions in a RF ion trap and DCE isomers under controlled conditions. Through systematic manipulation of the Ca$^+$ electronic state populations \textit{via} tuning of laser cooling parameters, we explore the interplay between molecular geometry and quantum-state effects in determining reaction outcomes and kinetics. Our experiments reveal a rich chemistry dominated by two primary reaction channels: formation of CaCl$^+$ and production of C$_2$H$_2$CaCl$^+$ which subsequently undergo secondary reactions in the trap. By combining state-resolved kinetic measurements with quantum chemical calculations, we elucidated reaction mechanisms that reveal pronounced reactivity differences not only between the geometric isomers of DCE, but also between the ground and excited states of Ca$^+$.

The present investigation provides insights into isomer-specific ion--molecule reaction dynamics while capitalising on the capabilities of Coulomb-crystal methodologies for probing state-selected reaction dynamics. The observed state- and isomer-dependent reactivities serve as benchmarks for theoretical models of ion--molecule reactions, contributing to our fundamental understanding of how molecular structure and quantum states govern chemical reactivity.

\section{Methods}

\subsection{Experimental methods}
The experimental setup employed in this study has been described previously~\cite{chang13a,roesch16a,kilaj18a,xu24a} such that we only briefly discuss the details relevant for the present work. Ca$^+$ ions were produced by non-resonant photoionisation of neutral calcium atoms emanating from a resistively heated calcium oven using a Nd:YAG laser (Quantel Brilliant, 355 nm, 5 ns) in the centre of a linear RF ion trap where they were subsequently confined. The trapped Ca$^+$ ions were then laser-cooled using radiation at 397 nm and 866 nm~\cite{willitsch17a}. By fixing the detuning and polarisation of the 866~nm laser while systematically adjusting the detuning of the 397~nm laser, we achieved precise control over the relative population of Ca$^+$ in the (4s)$^2$S$_{1/2}$, (4p)$^2$P$_{1/2}$ and (3d)$^2$D$_{3/2}$ states involved in the laser-cooling cycle. The laser frequencies were continuously monitored with a wavemeter (HighFinesse WSU-30) calibrated by a frequency-stabilised HeNe reference laser (Thorlabs HRS015). In the present study, the laser wavelength was detuned by 10, 20, and 30~fm from the atomic resonance, corresponding to frequency detunings of $-19$, $-38$, and $-57$~MHz, respectively. 

The neutral reactants, \textit{trans}- and \textit{cis}-1,2-DCE (Thermo Fisher Scientific, $99\%$ and $97\%$ purity, respectively), were contained in separate metal vessels connected to a leak valve (Lesker, LVM940). To ensure sample purity in the gas system, the relevant vessel was immersed in an ethanol--dry ice bath for five minutes to freeze the liquid sample, followed by evacuation of residual gas from the container. After allowing the sample to equilibrate to room temperature, it was introduced into the reaction chamber through the leak valve at a nominal pressure of 1.00 $\times$ 10$^{-9}$~mbar (with a 5\% uncertainty from fluctuations of the gauge reading), as measured by a Pfeiffer compact cold-cathode gauge (IKR 270) mounted above the ion trap. To determine the partial pressure of DCE in the reaction chamber, the molecule-specific detection efficiency of the pressure gauge was taken into account~\cite{bartmess83a,nakao75a}. Our gauge calibration was based on previously reported comparative measurements of various gases~\cite{bartmess83a}. While gauge correction factors specific for DCE do not seem to have been reported in the literature, we adopted a value of 4.53 reported for tetrachloroethylene given the structural similarity and similar ionisation energy with DCE~\cite{haynes16a}. Applying this correction to our measured pressure yielded an actual DCE pressure of 0.22 × 10$^{-9}$ mbar in the reaction chamber (with an estimated uncertainty of a factor of 2, see Section 3 of the Supplementary Information (SI) for details). Previous studies have established an isomerisation barrier of approximately 2.4~eV between \textit{cis}- and \textit{trans}-DCE~\cite{jeffers72a}, which precludes thermal isomerisation under our experimental conditions. Pseudo-first-order reaction rate coefficients $k_{\text{pseudo}}$ for the reactions of Ca$^+$ with the \textit{cis} and \textit{trans} isomers of DCE were determined at different detunings from resonance of the 397~nm laser. For measurements of the kinetics and the quantitative analysis of the reaction products, ions were ejected from the trap into a time-of-flight mass spectrometer (TOF-MS) after well-defined reaction periods. Bimolecular rate coefficients were obtained as $k = k_{\text{pseudo}}/n_{\text{DCE}}$ assuming a constant number density $n_{\text{DCE}} = 5.3 \times 10^6$ cm$^{-3}$ for DCE in the reaction chamber at a temperature of 300~K calculated from the corrected pressure-gauge reading. 

\subsection{Capture theory predictions}

To interpret the observed isomer-specific reactivities, we compared our experimental results with predictions from capture-theory models. For non-polar species such as \textit{trans}-DCE, the Langevin model~\cite{gioumousis58a,tsikritea22a} provides a suitable framework. In this approach, the rate coefficient is given by

\begin{equation} k_L = q \sqrt{\frac{\pi \alpha'}{\epsilon_0 \mu}}, \end{equation}

where \(q\) denotes the ionic charge, \(\alpha'\) is the polarizability volume of the neutral species (determined previously to be $\alpha' = 8.15 \times 10^{-24}$~cm$^3$ for \textit{trans}-DCE~\cite{haynes16a}), \(\epsilon_0\) is the vacuum permittivity, and \(\mu\) is the reduced mass of the ion--molecule pair.

By contrast, for polar species such as \textit{cis}-DCE, Average Dipole Orientation (ADO) theory~\cite{su73b} offers a more suitable treatment by explicitly incorporating ion--dipole interactions. According to the ADO model, the rate coefficient is given by

\begin{equation} k_{\text{ADO}} = q \sqrt{\frac{\pi \alpha'}{\epsilon_0 \mu}} + \frac{q \mu_D c}{\epsilon_0} \sqrt{\frac{1}{2 \pi \mu k_B T}}. \end{equation}

Here, in addition to the parameters defined above, $\mu_D$ is the dipole moment of the neutral species, $k_B$ is the Boltzmann constant, $T$ is the temperature, and $c$ is an empirical orientation parameter that quantifies the effectiveness of ion--dipole alignment and is usually determined through experimental benchmarking.

Previous experimental work~\cite{rebrion88b} on N$^+$ and H$_3^+$ reactions with DCE provides a useful reference point for determining this parameter. Analysis of their measured rate coefficients under comparable experimental conditions shows good agreement with ADO theory predictions when using an orientation parameter of $c \approx 0.14$ for \textit{cis}-DCE (see Section 2 of the SI). Given the similar nature of our reaction conditions, we adopt this value for our current investigation.

\subsection{Quantum-chemical calculations}

While the analysis of long-range interactions can provide insights into reaction kinetics, a comprehensive understanding of the reaction mechanism and product formation requires investigation of short-range effects. To this end, we performed quantum-chemical calculations of the potential energy surface (PES) of the present reaction system. Initial geometry optimisations of reactants, products, transition states, and reaction intermediates were performed at the UPBE0~\cite{adamo99a}/6-311++g(d,p) density-functional-theory (DFT) level with GD3BJ dispersion corrections~\cite{grimme11a,becke05a} to account for van-der-Waals interactions in the Ca$^+$--DCE system. At the same level of theory, we computed intrinsic reaction coordinates (IRCs) to verify reaction pathways, along with zero-point energy (ZPE) corrections.

To achieve higher accuracy in electronic energies and barrier heights, single-point energy calculations were performed at the UCCSD(T)~\cite{raghavachari89a}/cc-pVQZ~\cite{dunning89a} level on the DFT-optimised structures. All calculations were performed using the Gaussian 16 software package~\cite{g16}.

\subsection{Optical Bloch equation modelling}

To quantify the electronic-state populations of the Ca\(^{+}\) ions, we recorded the fluorescence associated with the $^2$P$_{1/2}\rightarrow\, ^2$S$_{1/2}$ transition while systematically varying the detuning of the 397~nm laser. In a second measurement series, the detuning of the 866~nm laser was varied.
The resulting fluorescence spectra were simultaneously fitted to an eight-level optical Bloch equation (OBE) model (see Section 4 of the SI). To ensure reliable parameter estimation, we employed a multi-start global optimisation approach~\cite{marti03a} in which 10000 local optimisations, each initialised with random guesses drawn from physically motivated bounds, were carried out. The resulting parameters of the global optimum were then used to simulate the population distributions among the $^2$S$_{1/2}$, $^2$P$_{1/2}$, and $^2$D$_{3/2}$ electronic states.

\begin{figure*}[!t]
\centering
\includegraphics[width=\textwidth]{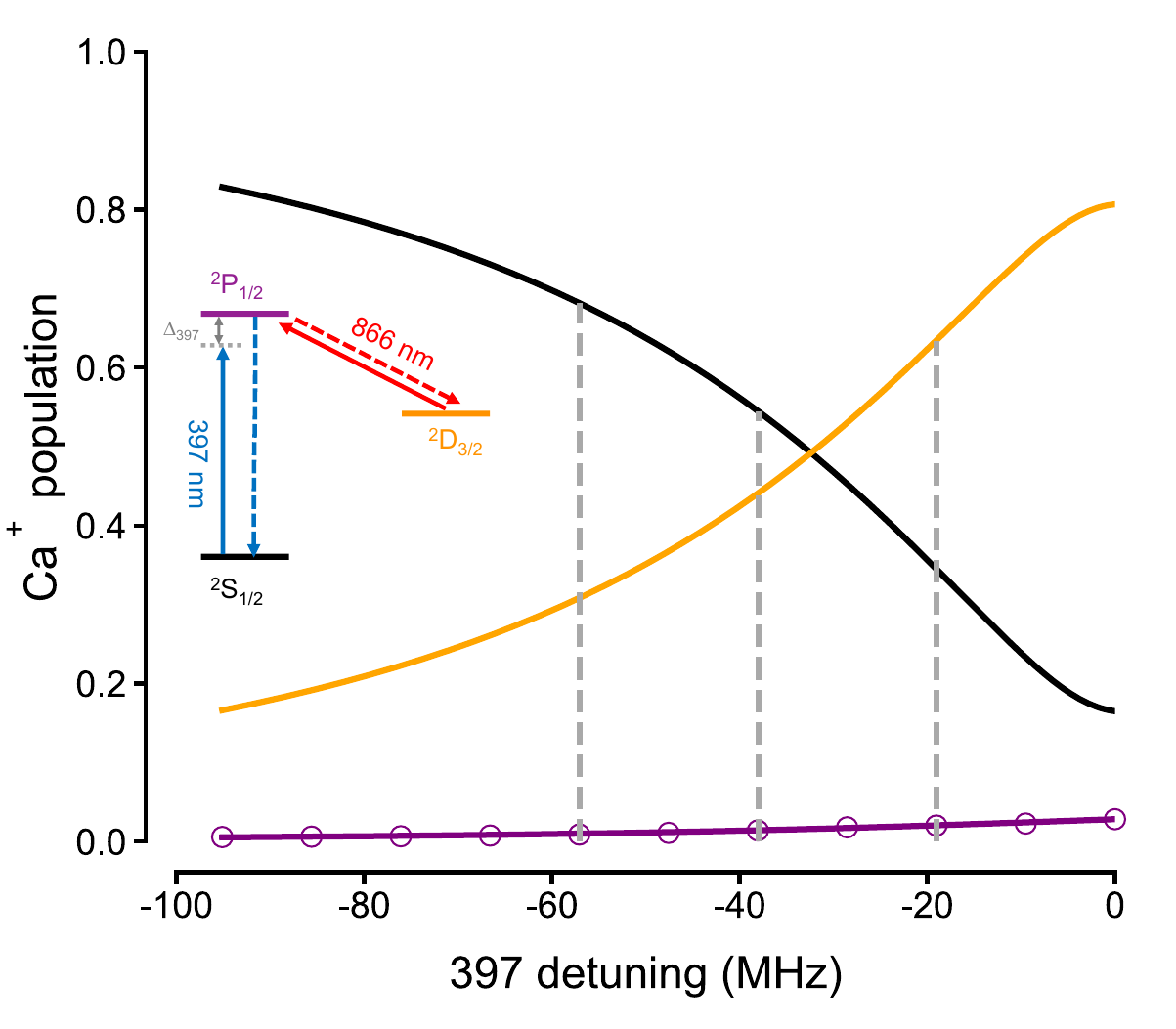}
\caption{Ca$^+$ electronic state population as a function of 397~nm cooling-laser detuning. Black, orange, and purple curves show the populations of the $^2$S$_{1/2}$, $^2$D$_{3/2}$, and $^2$P$_{1/2}$ states obtained from optical Bloch equation (OBE) simulations. Open circles represent experimental fluorescence data scaled to match the simulations. The inset illustrates the laser cooling scheme with 397 nm and 866 nm lasers coupling the three electronic states. Dashed lines indicate spontaneous emission channels.}
\label{fig:state_dependent}
\end{figure*}

Figure~\ref{fig:state_dependent} shows the simulated state populations as a function of 397~nm laser detuning obtained from the OBE simulations. The simulation reveals that as the detuning increases (moving closer to resonance), the total population of the excited states ($^2$P$_{1/2}$ + $^2$D$_{3/2}$) increases, while the ground-state $^2$S$_{1/2}$ population correspondingly decreases. For the kinetic analysis presented in this work, we distinguish between the ground state ($^2$S$_{1/2}$) and the combined excited-state population ($^2$P$_{1/2}$ + $^2$D$_{3/2}$), rather than treating each excited state separately. These population distributions provide the basis for extracting ground and excited state rate coefficients from our experimental measurements (see Section~\ref{sec:state_kinetics}).

\section{Results}
\subsection{Reaction products}

\begin{figure*}[!t]
\centering
\includegraphics[width=\textwidth]{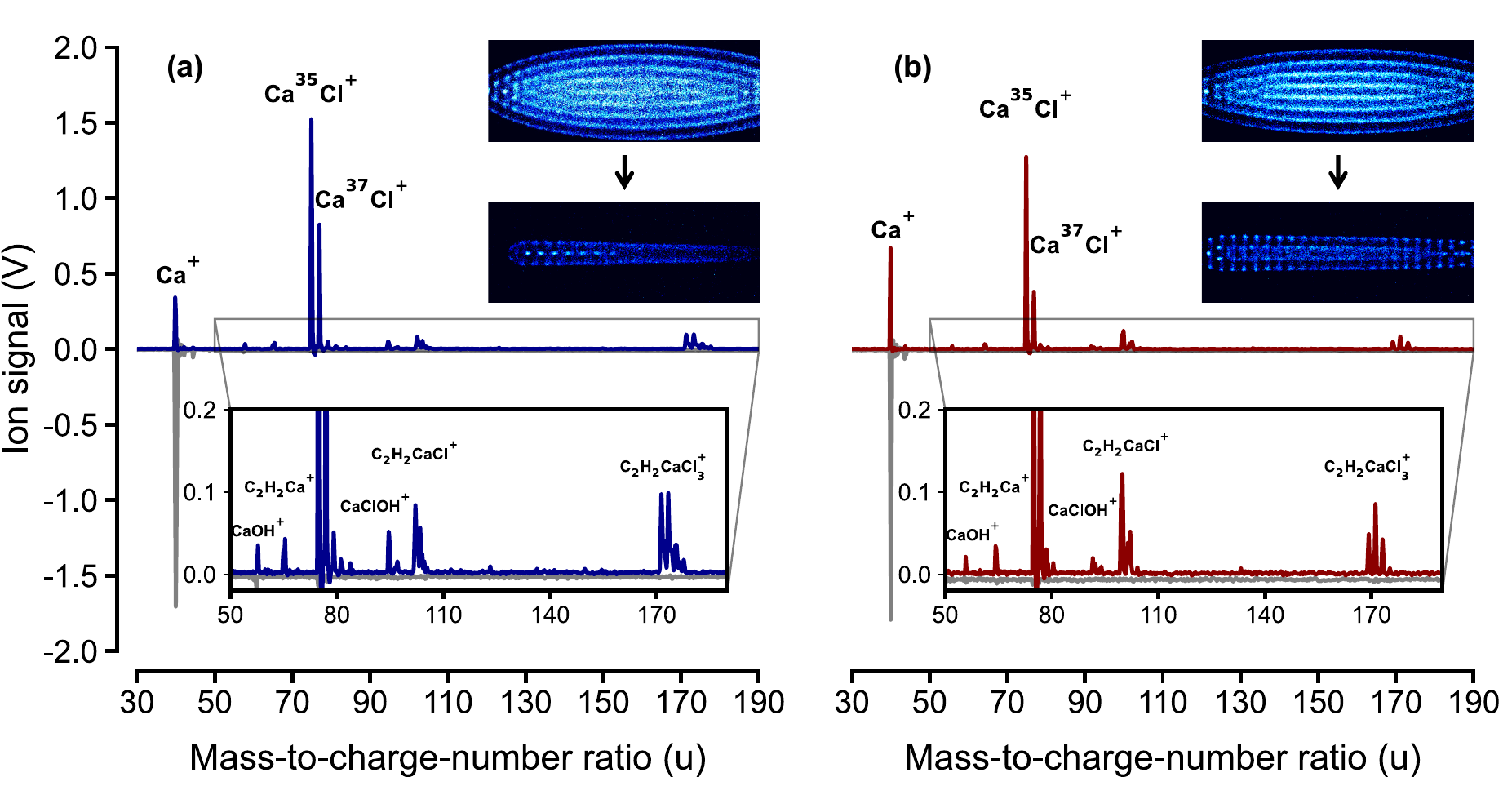}
\caption{Time-of-flight mass spectra of the reactions of Ca$^+$ Coulomb crystals with (a) \textit{cis}- and (b) \textit{trans}-DCE after 4 minutes of reaction time compared with background spectra taken without exposure to DCE gas (grey inverted traces). Top insets:  fluorescence images of the Coulomb crystal before and after 4 minutes of reaction with DCE.  Bottom insets: magnification of the mass spectra in the range $m/z = 50-190$~u to highlight product ions. Each spectrum represents an average of 9 individual measurements.}
\label{fig:mass_spectra}
\end{figure*}

Figure \ref{fig:mass_spectra} shows representative TOF mass spectra obtained after 4 minutes of reaction time at 10~fm ($-19$~MHz) detuning of the 397~nm cooling laser for both \emph{cis-} (blue traces) and \emph{trans-}DCE (red traces). The observed peaks can be assigned based on their mass-to-charge-number ratios ($m/z$) and characteristic isotopic patterns: CaOH$^+$ ($m/z = 57$~u), C$_2$H$_2$Ca$^+$ ($m/z = 66$~u), CaCl$^+$ ($m/z = 75\,$u and $ 77\,$u, showing the characteristic $\approx 3:1$ isotope ratio of $^{35}$Cl/$^{37}$Cl), C$_2$H$_2$CaCl$^+$ ($m/z = 101\,$u and $103\,$u), and C$_2$H$_2$CaCl$_3^+$ ($m/z = 171\,$u$, 173\,$u$, 175\,$u, and $177\,$u, exhibiting a characteristic isotopic intensity pattern caused by three chlorine atoms). Additional peaks at $m/z = 92\,$u and $94\,$u were assigned to CaClOH$^+$. Note that due to the inherent mass-calibration limitations of our TOF-MS, ions heavier than Ca$^+$ exhibit uncertainties in the mass assignment of up to $\pm$2 u.
Our results reveal that CaCl$^+$ and C$_2$H$_2$CaCl$^+$ formation represent the predominant reaction channels.
A particularly noteworthy aspect is the observation of a secondary reaction wherein C$_2$H$_2$CaCl$^+$ subsequently reacts with additional DCE to form C$_2$H$_2$CaCl$_3^+$.

The minor species detected in the TOF mass spectra provide additional insights into the present reaction network. CaOH$^+$ likely forms through reactions of Ca$^+$ with the residual water in the vacuum chamber~\cite{okada02a,wu24a}, while CaClOH$^+$ may originate from secondary reactions of CaCl$^+$ with the same background gas. The trace amounts of C$_2$H$_2$Ca$^+$ and C$_2$HCaCl$_2^+$ ($m/z = 135\,$u$, 137\,$u, and $139\,$u, barely detectable in the TOF-MS) represent minor reaction channels, similar to previous reports~\cite{mikhailov06a, bodi11a}. Previous photoelectron-photoion coincidence (PEPICO) studies of DCE~\cite{bodi11a} revealed that C$_2$H$_2^+$ (Cl$_2$ elimination) and C$_2$HCl$_2^+$ (H elimination) ions exhibit higher appearance energies compared to the Cl elimination channel. This observation suggests that the formation of these minor ionic products observed in our work likely proceeds through reaction with excited electronic states of Ca$^+$, rather than through the ground state. Given their small contribution to the overall product yield ($<$2\%), these minor channels were excluded from our kinetic analysis.

\subsection{Isomer-specific reaction kinetics}
To characterise the reaction kinetics, we analysed the temporal evolution of product ions by integrating the area under the different peaks in the mass spectra at various reaction times. This time-resolved analysis allowed us to track the formation and consumption of various ionic species throughout the reaction process.
Based on the observed products, we established a kinetic model comprising three main reaction pathways:
\begin{align*} 
    \text{Ca}^+ + \text{C}_2\text{H}_2\text{Cl}_2 & \xrightarrow{k_1} \text{CaCl}^+ + \text{C}_2\text{H}_2\text{Cl} \tag{Reaction 1} \\ 
    \text{Ca}^+ + \text{C}_2\text{H}_2\text{Cl}_2 & \xrightarrow{k_2} \text{C}_2\text{H}_2\text{CaCl}^+ + \text{Cl} \tag{Reaction 2} \\ 
    \text{C}_2\text{H}_2\text{CaCl}^+ + \text{C}_2\text{H}_2\text{Cl}_2 & \xrightarrow{k_3} \text{C}_2\text{H}_2\text{CaCl}_3^+ + \text{C}_2\text{H}_2 \tag{Reaction 3}  
\end{align*} 

The reaction mechanism consists of two competing primary channels characterised by bimolecular rate coefficients $k_1$ and $k_2$, leading to the formation of CaCl$^+$ and C$_2$H$_2$CaCl$^+$, respectively. A subsequent secondary reaction pathway proceeds with rate coefficient $k_3$, where C$_2$H$_2$CaCl$^+$ further reacts with DCE to produce C$_2$H$_2$CaCl$_3^+$. The temporal evolution of these ionic species can be described by the rate equations:

\begin{align}
    \frac{d[\text{Ca}^+]}{dt} &= -k_{\text{tot}}[\text{Ca}^+][\text{DCE}] \label{eq:ca} \\
    \frac{d[\text{CaCl}^+]}{dt} &= k_1[\text{Ca}^+][\text{DCE}] \label{eq:cacl} \\
    \frac{d[\text{C}_2\text{H}_2\text{CaCl}^+]}{dt} &= k_2[\text{Ca}^+][\text{DCE}] - k_3[\text{C}_2\text{H}_2\text{CaCl}^+][\text{DCE}] \label{eq:c2hcacl} \\
    \frac{d[\text{C}_2\text{H}_2\text{CaCl}_3^+]}{dt} &= k_3[\text{C}_2\text{H}_2\text{CaCl}^+][\text{DCE}] \label{eq:c2hcacl3}
\end{align}

The total rate coefficient $k_{\text{tot}}$ represents the sum of the primary reaction channels ($k_{\text{tot}} = k_1 + k_2$).

\begin{figure*}[ht]
\centering
\includegraphics[width=\textwidth,trim={0 0cm 0 0cm},clip]{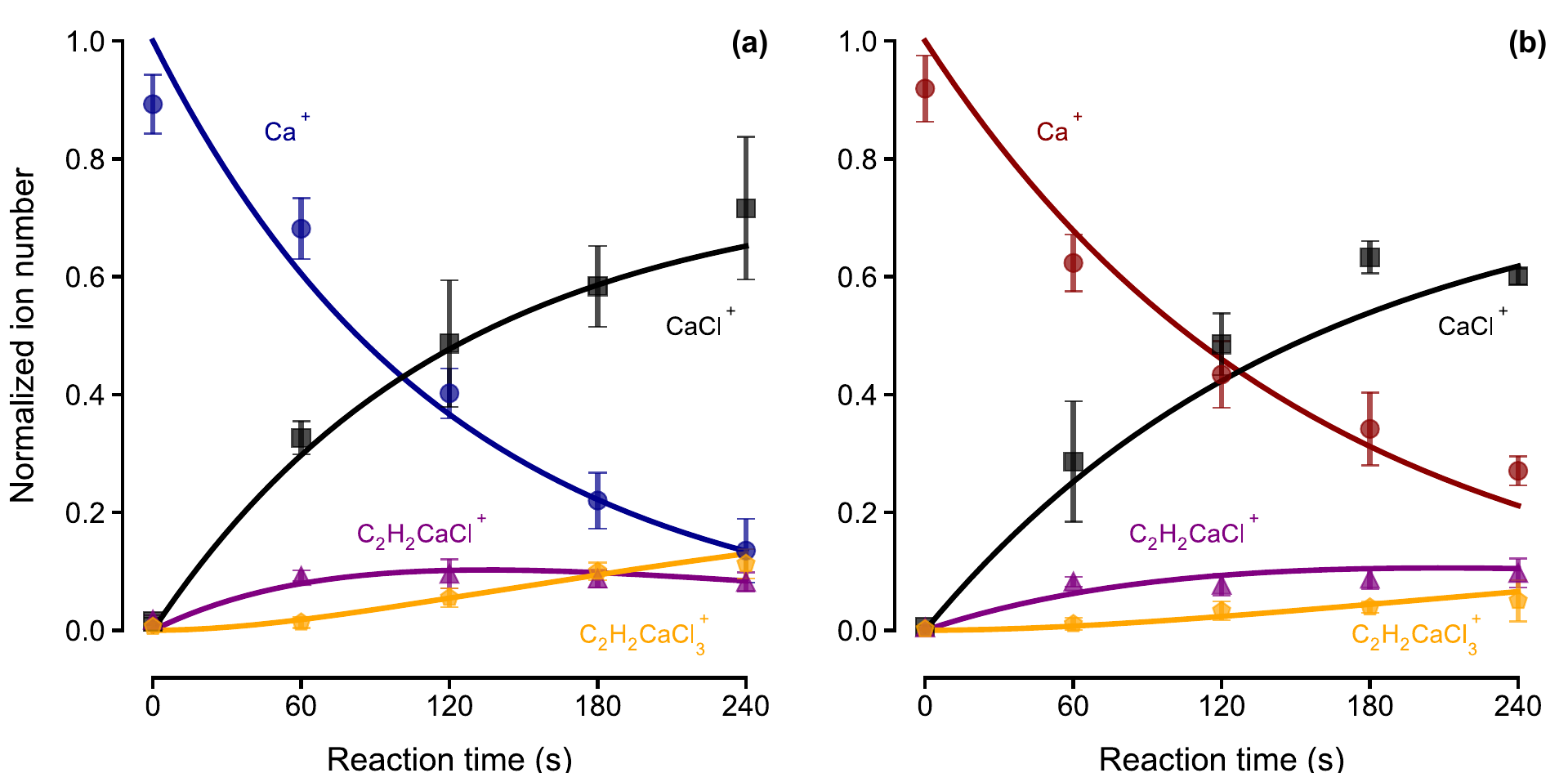}
\caption{Representative reaction kinetics of Ca$^+$ reacting with (a) \textit{cis}-DCE and (b) \textit{trans}-DCE at a 10~fm detuning of the 397~nm cooling laser. The blue and red curves represent the relative decay of the number of Ca$^+$ ions for the \textit{cis} and \textit{trans} reactions, respectively, while the emergence of CaCl$^+$ (black), C$_2$H$_2$CaCl$^+$ (purple), and C$_2$H$_2$CaCl$_3^+$ (orange) reflects the progression of primary and secondary reaction pathways. All data points are normalised using a scaling factor $S$ obtained from kinetic modelling (see SI for details). Solid lines denote fits to the rate-equation model Equations (\ref{eq:ca})-(\ref{eq:c2hcacl3}) in the main text. Error bars represent standard deviations calculated from three independent experimental measurements.}
\label{fig:kinetics}
\end{figure*}

Figure \ref{fig:kinetics} illustrates the reaction kinetics for both \textit{cis}- and \textit{trans}-DCE reacting with Ca$^+$ at a 10~fm detuning of the 397~nm cooling laser. The data shown represent one complete measurement from a series of three independent experimental runs performed in separate experimental sessions. The simultaneous appearance of C$_2$H$_2$CaCl$^+$ and CaCl$^+$ confirms their formation through competing primary reaction channels. The delayed formation of C$_2$H$_2$CaCl$_3^+$ and its kinetic profile suggests its origin as a secondary reaction product from C$_2$H$_2$CaCl$^+$ with DCE.

The kinetic data presented in this study were collected using a systematic experimental protocol to ensure reliability and reproducibility. For each laser detuning, we performed a total of nine measurements organised as follows: three independent experimental sessions were conducted, with each session comprising three consecutive replicate measurements. For the data analysis, we first calculated the reaction-rate coefficient for each experimental session by fitting the three replicate measurements together. This yielded three independent values for the rate coefficients per detuning (one from each session). The final rate coefficients reported throughout this paper represent the mean of these three session-specific values, with uncertainties expressed as standard error between sessions, thereby capturing the experimental reproducibility across different measurement days.

To analyse the reaction kinetics, we employed a fitting procedure where the parameters of our rate equation model were adjusted to reproduce the temporal evolution of the integrated ion signals. All integrated signals were normalised using a scaling factor $S$ obtained from the fitting, as shown in Figure~\ref{fig:kinetics}. This approach ensured consistent initial conditions for our kinetic analysis (see Section 1 of the SI for detailed methodology).

Table~\ref{tbl:rate_constants} presents the bimolecular rate coefficients obtained by fitting the experimental kinetic data at various cooling-laser detunings. The formation rate of CaCl$^+$ ($k_1$) seems to exhibit a slight inverse dependence on the detuning, decreasing from 1.19 × 10$^{-9}$ to 0.90 × 10$^{-9}$ cm$^3$s$^{-1}$ for \textit{cis}-DCE and from 0.95 × 10$^{-9}$ to 0.74 × 10$^{-9}$ cm$^3$s$^{-1}$ for \textit{trans}-DCE as the detuning increases from 10 to 30 fm. By contrast, the formation rate of C$_2$H$_2$CaCl$^+$ ($k_2$) appears nearly constant across all considered detunings for both isomers. \textit{cis}-DCE consistently shows approximately 25\% higher reactivity than the \textit{trans} isomer, with this ratio remaining stable within the experimental uncertainties. Moreover, the branching ratio ($k_1/k_2$) seems to decrease with increasing detuning from 3.5 to 2.4 for the \textit{cis} isomer and from 3.6 to 2.8 for the \textit{trans} isomer.

\begin{table}[H]
\small
 \caption{Averaged bimolecular rate coefficients ($10^{-9}$ cm$^3$s$^{-1}$) for different reaction channels at the indicated detunings of the cooling laser from resonance. Values in parentheses represent the standard error of the mean of three independent measurements. }
  \label{tbl:rate_constants}
  \begin{tabular*}{\columnwidth}{@{\extracolsep{\fill}}lccc}
    \hline\hline
    Reaction Channel & 10 fm & 20 fm & 30 fm \\
    \hline
    Total rate coefficient ($k_{\text{tot}}$) & & & \\
    \textit{cis} & 1.5(1) & 1.4(2) & 1.3(1) \\
    \textit{trans} & 1.21(3) & 1.07(2) & 1.00(4) \\
    \hline
    CaCl$^+$ formation ($k_1$) & & & \\
    \textit{cis} & 1.19(9) & 1.1(2) & 0.90(9) \\
    \textit{trans} & 0.95(5) & 0.81(3) & 0.74(3) \\
    \hline
    C$_2$H$_2$CaCl$^+$ formation ($k_2$) & & & \\
    \textit{cis} & 0.34(2) & 0.30(1) & 0.37(2) \\
    \textit{trans} & 0.27(2) & 0.26(1) & 0.27(1) \\
    \hline
    C$_2$H$_2$CaCl$_3^+$ formation ($k_3$) & & & \\
    \textit{cis} & 1.2(2) & 0.89(9) & 1.1(3) \\
    \textit{trans} & 0.57(4) & 0.47(9) & 0.30(6) \\
    \hline
    Branching ratio ($k_1/k_2$) & & & \\
    \textit{cis} & 3.5(3) & 3.6(5) & 2.4(3) \\
    \textit{trans} & 3.6(3) & 3.1(2) & 2.8(2) \\
    \hline\hline
  \end{tabular*}
\end{table}

\subsection{State-specific reaction kinetics}
\label{sec:state_kinetics}
These results suggest that the reaction kinetics are influenced by the laser detuning, which directly controls the populations in the (4s)$^2$S$_{1/2}$, (4p)$^2$P$_{1/2}$ and (3d)$^2$D$_{3/2}$ electronic states of Ca$^+$. At each specific laser detuning, the populations maintain a dynamic equilibrium remaining constant throughout the reaction. Therefore, the rate coefficients reported in the previous section represent averaged values weighted by the populations of these electronic states.

\begin{figure*}[!h]
\centering
\includegraphics[width=\textwidth]{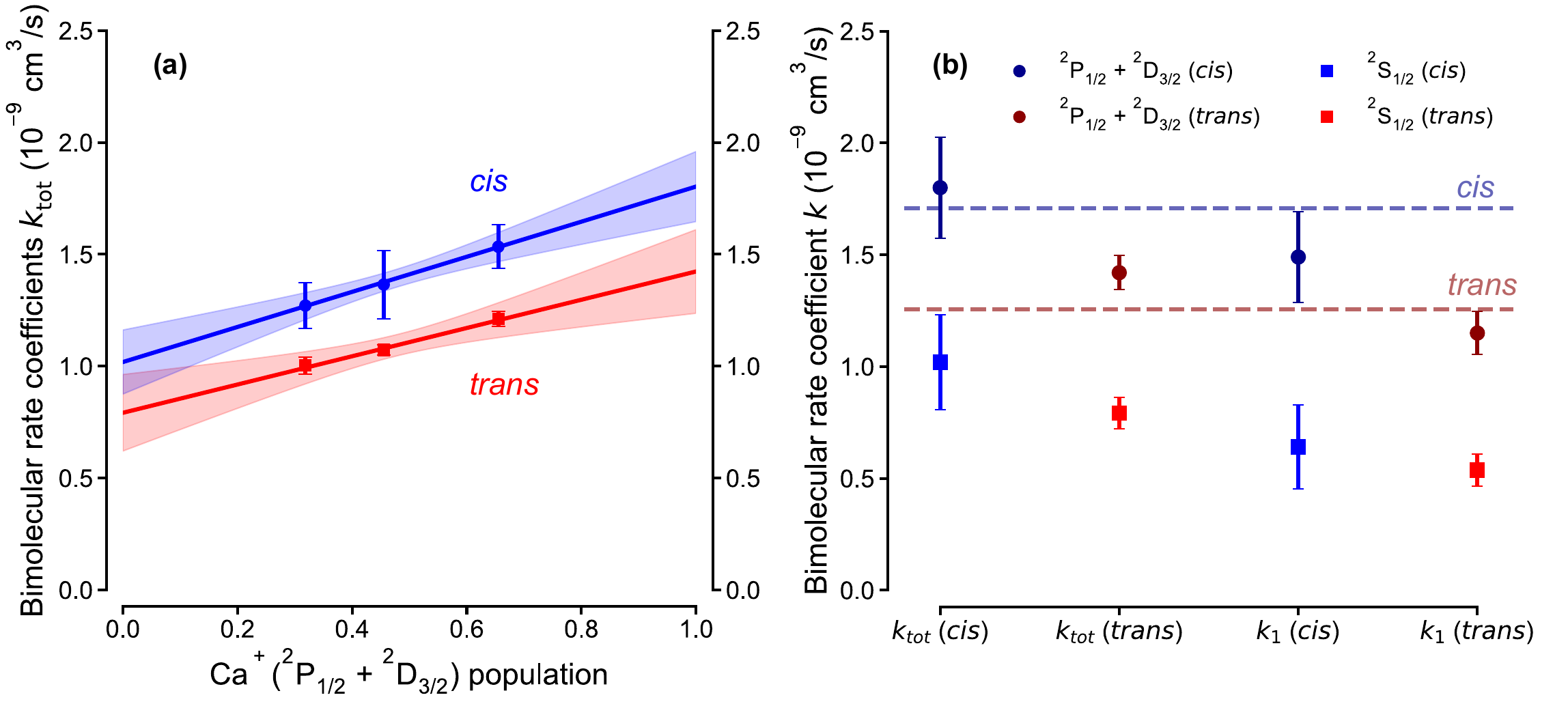}
\caption{(a) Total bimolecular rate coefficients ($k_{\text{tot}}$) as a function of the combined excited state ($^2$P$_{1/2}$ + $^2$D$_{3/2}$) population fraction. Blue and red points represent experimental data for \textit{cis}- and \textit{trans}-DCE, respectively, with error bars showing the standard error of three independent measurements. Solid lines show linear fits to the data with shaded areas indicating 90\% confidence regions. (b) Comparison of bimolecular rate coefficients for reactions of Ca$^+$ in different electronic states with \textit{cis}- and \textit{trans}-DCE. Blue symbols represent data for \textit{cis}-DCE, and red symbols for \textit{trans}-DCE. Squares indicate reactions with ground state Ca$^+$ ($^2$S$_{1/2}$), while circles denote reactions with excited state Ca$^+$ ($^2$P$_{1/2}$ + $^2$D$_{3/2}$). Horizontal lines represent the ADO and Langevin-theoretical predictions for \textit{cis}-DCE (blue dashed line) and \textit{trans}-DCE (red dashed line), respectively. Error bars represent the standard error of the fitted state-specific rate coefficients.}
\label{fig:state_dependent_rates}
\end{figure*}

As discussed in Section 2.4 and shown in Figure~\ref{fig:state_dependent}, the OBE simulations provide the electronic state populations as a function of laser detuning. To extract state-specific rate coefficients, we fitted a linear model $k^{c/t}_{x} = (1 - p_x)k^{c/t}_{\text{S}} + p_xk^{c/t}_{\text{P+D}}$ to the measured rate coefficients $k^{c/t}_{x}$ as a function of the combined $^2$P$_{1/2}$ + $^2$D$_{3/2}$ excited-state population $p_x$, as shown in Figure~\ref{fig:state_dependent_rates}(a). For the total rate coefficients of \textit{cis}-DCE, we obtain $k^c_{\text{tot,S}} = 1.0(2) \times 10^{-9}$ cm$^3$s$^{-1}$ for ground state Ca$^+$ and $k^c_{\text{tot,P+D}} = 1.8(2) \times 10^{-9}$ cm$^3$s$^{-1}$ for excited states. Similarly, for \textit{trans}-DCE, we obtained $k^t_{\text{tot,S}} = 0.79(7) \times 10^{-9}$ cm$^3$s$^{-1}$ and $k^t_{\text{tot,P+D}} = 1.4(1) \times 10^{-9}$ cm$^3$s$^{-1}$.  Examining specifically the CaCl$^+$ formation channel ($k_1$), \textit{cis}-DCE exhibits rate coefficients of $k^c_{1,\text{S}} = 0.6(2) \times 10^{-9}$ cm$^3$s$^{-1}$ and $k^c_{1,\text{P+D}} = 1.5(2) \times 10^{-9}$ cm$^3$s$^{-1}$, while \textit{trans}-DCE shows $k^t_{1,\text{S}} = 0.54(7) \times 10^{-9}$ cm$^3$s$^{-1}$ and $k^t_{1,\text{P+D}} = 1.2(1) \times 10^{-9}$ cm$^3$s$^{-1}$. These results are plotted in Figure~\ref{fig:state_dependent_rates} (b).

The capture theory predicts rate coefficients of $1.7 \times 10^{-9}$ cm$^3$s$^{-1}$ and $1.3 \times 10^{-9}$ cm$^3$s$^{-1}$ for \textit{cis}- and \textit{trans}-DCE, respectively, shown as horizontal lines in Figure~\ref{fig:state_dependent_rates} (b). Our experimental results for excited-state Ca$^+$ agree with these theoretical predictions. The rate coefficients for the ground-state reactions deviate from the capture-theory predictions by about a factor of 2, suggesting a common mechanistic origin affecting both isomers.

\subsection{Quantum-chemical calculations}

\begin{figure*}[htbp]
\centering
\includegraphics[width=\textwidth]{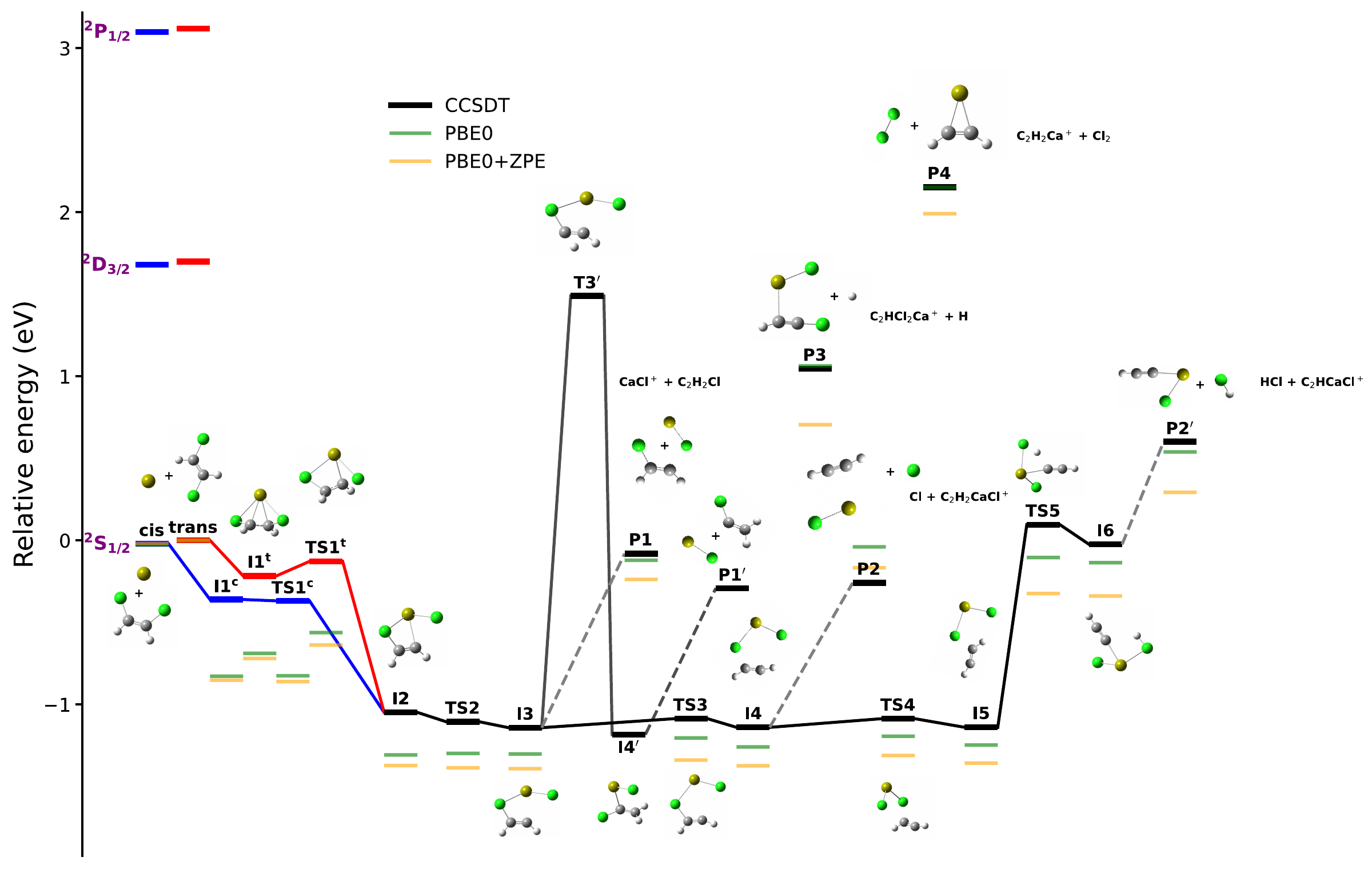}
\caption{Calculated potential energy surfaces for the reactions of ground state Ca$^+$($^2$S$_{1/2}$) with \textit{cis}- and \textit{trans}-DCE. All molecular structures were optimised at the PBE0/6-311++g(d,p) level of theory, with single-point energies calculated at the UCCSD(T)/cc-pVQZ level. For comparison, energies at the PBE0/6-311++g(d,p) level without (green lines) and with (orange lines) zero-point energy corrections are also shown. The blue and red pathways represent the initial reaction pathways for \textit{cis}- and \textit{trans}-DCE respectively, which converge to a common pathway (black) after isomerisation. The horizontal lines at 0.00, 1.70, and 3.12 eV indicate the energies of the $^2$S$_{1/2}$, $^2$D$_{3/2}$, and $^2$P$_{1/2}$ electronic states of Ca$^+$ corresponding to the available entrance channels for the reactions. Molecular geometries of key intermediates and transition states are shown, with Cl atoms in green, Ca in yellow, C in grey, and H in white. All energies are given in eV relative to the \textit{trans}-DCE + Ca$^+$($^2$S$_{1/2}$) reactant asymptote.}
\label{fig:DCE_Ca_PES}
\end{figure*}

These experimental results provide initial insights into the reaction mechanisms, but a deeper understanding requires quantum-chemical calculations to interpret and rationalise the observations. Figure~\ref{fig:DCE_Ca_PES} shows relevant stationary points on the potential-energy surface of the present reaction system connecting the reactants to the observed products. Our quantum chemical calculations reveal that \textit{cis}-DCE possesses slightly lower energy (0.02~eV) than \textit{trans}-DCE, contrary to conventional chemical intuition. While this computed energy difference falls within the typical accuracy limit of CCSD(T) calculations (chemical accuracy of $\sim$1~kcal/mol), the energetic ordering ($E_{\mathrm{cis}} < E_{\mathrm{trans}}$) is well established through independent evidence. This counterintuitive energy ordering is a manifestation of the ``\textit{cis} effect'', a phenomenon where the \textit{cis} isomer is more stable than its \textit{trans} counterpart, in 1,2-dihaloethylene molecules and is consistent with previous experimental isomerisation studies~\cite{jeffers72a, craig71a} and theoretical calculations~\cite{kanakaraju02a} of DCE. While Figure~\ref{fig:DCE_Ca_PES} shows calculations of ground-state energies, we have also indicated the energies of the asymptotes corresponding to the excited $^2$P$_{1/2}$ and $^2$D$_{3/2}$ states of Ca$^+$ in the entrance channel, which lie 1.70 and 3.12 eV above the ground state, respectively.

The reaction is initiated by the formation of Ca$^+$--DCE complexes in both isomers yielding intermediates I1$^{\text{c}}$ and I1$^{\text{t}}$ for \textit{cis}- and \textit{trans}-DCE, respectively. Subsequently, the \textit{cis}-DCE complex follows a barrierless path through intermediates I2 to I3, ultimately dissociating into CaCl$^+$ and C$_2$H$_2$Cl (P1). In contrast, the reaction dynamics of the \textit{trans} isomer reveals an intriguing feature: despite the significant \emph{cis/trans-}isomerisation energy (2.4 eV) determined for neutral DCE~\cite{jeffers72a}, the presence of Ca$^+$ seems to substantially lower this barrier, enabling the \textit{trans}-DCE complex to undergo isomerisation \textit{via} TS1$^{\text{t}}$ to access the same intermediate I2 as the \textit{cis} pathway. Following this Ca$^+$-mediated isomerisation, both pathways converge and proceed through an identical route to the products (black traces). 

The formation of C$_2$H$_2$CaCl$^+$ and Cl (P2) represents the energetically favoured chlorine elimination channel. Starting from intermediate I3, the reaction complex undergoes planarisation reaching intermediate I4. From I4, direct dissociation of a chlorine atom yields P2. P2 is a strongly bound electrostatic complex between CaCl$^+$ and C$_2$H$_2$ with a calculated binding energy of $\sim$0.9 eV at the UCCSD(T)/cc-pVQZ level, which lies approximately 0.26 eV below the reactant asymptote. This renders this channel exothermic and readily accessible for reactions with ground-state Ca$^+$, and the product stable under the current experimental conditions.

An alternative reaction pathway involves hydrogen migration leading to HCl elimination (P2$'$). This pathway proceeds through a torsional motion about the Cl-C bond from I4 to form intermediate I5, followed by hydrogen transfer to a chlorine atom \textit{via} transition state TS5, forming the complex I6 before dissociating into C$_2$HCaCl$^+$ and HCl (P2$'$). Notably, the hydrogen migration process requires overcoming a significant energy barrier of 1.24 eV, and the final products P2$'$ are approximately 0.6 eV higher in energy than the reactants. These energetics suggest that the HCl elimination channel is not accessible for the reactions of ground-state Ca$^+$ with DCE, given that the average collision energy is only 0.04 eV, which is insufficient to overcome the endothermicity. 

The energetics of the trace products C$_2$H$_2$Ca$^+$ and C$_2$HCaCl$_2^+$ observed in the experiments were also computed and are indicated in Figure~\ref{fig:DCE_Ca_PES}. Our calculations reveal that the H-loss channel leading to C$_2$HCaCl$_2^+$ (P3) becomes accessible when Ca$^+$ is in the excited $^2$P$_{1/2}$ or $^2$D$_{3/2}$ states, while the Cl$_2$-loss channel yielding C$_2$H$_2$Ca$^+$ (P4) is only open when Ca$^+$ is in the $^2$P$_{1/2}$ state. This energetic hierarchy parallels the fragmentation behaviour observed in iPEPICO studies of DCE cations~\cite{bodi11a}, where C$_2$HCl$_2^+$ fragments appear at lower photon energies compared to C$_2$H$_2^+$ fragments. Despite the different underlying processes—unimolecular dissociation in iPEPICO versus ion--molecule reaction in our system—the relative energetic ordering of these fragmentation pathways is preserved. Similar to these energetic considerations, our experimental results show that the yield of C$_2$H$_2$Ca$^+$ is higher than that of C$_2$HCaCl$_2^+$, with the latter featuring only weakly in our mass spectra.

\section{Discussion}

The theoretical results indicate that both \textit{cis}- and \textit{trans}-DCE converge to identical reaction pathways and product ions, a behaviour previously observed in ion--molecule reactions involving DCE~\cite{mikhailov06a,rebrion88b}. The analysis of the rate coefficients as a function of laser detuning revealed several notable characteristics. The most striking feature is that both isomers display nearly parallel trends in their total reaction rates, as can be seen from Table \ref{tbl:rate_constants}, maintaining a consistent \textit{cis}/\textit{trans} reactivity ratio of approximately 1.3 across all detunings. Although this ratio is slightly lower than the capture theory prediction of 1.4 (which, however, critically hinges on the assumed value of the ADO orientation parameter), it remains consistent with the present theoretical value within the experimental uncertainties. The similar dependence of both isomers on the Ca$^+$ electronic-state populations, evidenced by their parallel rate trends, strongly suggests that the rates are predominantly governed by long-range interactions. 

However, capture theory, while valuable for predicting overall reaction rates, has inherent limitations in describing state-specific reactions and their product branching ratios, which are primarily governed by short-range interactions. Analysis of the state-specific rate coefficients reveals a discrepancy between theoretical predictions and experimental observations, as we can see in Figure Figure~\ref{fig:state_dependent_rates} (b). While excited-state Ca$^+$ reactions yielding CaCl$^+$ proceed at rates ($k^{c/t}_{1,\text{P+D}}$) that align well with capture-theory predictions, the corresponding ground-state rate coefficients ($k^{c/t}_{1,\text{S}}$) are lower, reaching only approximately half of the predicted values. This reduction in ground-state reactivity is puzzling given that our quantum-chemical calculations indicate an effectively barrierless pathway for CaCl$^+$ formation from the Ca$^+$($^2$S$_{1/2}$) ground state. Similar examples of reduced ground-state reactivity have been observed in other ion--molecule reactions, for example, in Be$^+$ + H$_2$O~\cite{yang18a} and Ca$^+$ with CH$_3$F and CH$_3$Cl~\cite{gingell10a},  where this phenomenon was generally attributed to small or submerged barriers along the reaction pathway. 
Note, however, that in the present case the product asymptote P1 has been calculated to lie only 0.14~eV below the reactants. We thus hypothesise that the reactivity in the ground-state channel is suppressed by the formation of a long-lived reaction complex in the deep potential well which is associated with the structures I2--I5 in figure~\ref{fig:DCE_Ca_PES} and bound by about 1~eV with respect to the products. As the reactant and product (P1) asymptotes are nearly degenerate, this complex can be expected to decay both back to the reactants as well as to product P1 with about equal probability which would explain the observed reduction of the formation rate of P1 by a factor of 2 with respect to the capture-theory prediction.

There is also an alternative pathway for CaCl$^+$ formation (P1$'$ in Figure~\ref{fig:DCE_Ca_PES}). Product P1$'$ is a structural isomer of P1 and possesses lower energy, making it thermodynamically more favourable. However, the formation of P1$'$ involves hydrogen migration from one carbon atom to another, which faces a substantial isomerisation barrier of 2.64~eV via transition state T3$'$. Overcoming such a high barrier is thermodynamically impossible on the ground-state surface under our experimental conditions. 
By contrast, when Ca$^+$ is electronically excited, the reaction energetics changes. Whether in this case the reactions proceed on excited surfaces or converts to the ground-state surface is impossible to assess in the absence of detailed dynamics calculations involving all relevant channels. However, the observed enhanced reactivity in the excited channels is consistent with more efficient pathways that potentially circumvent the energetic or geometric constraints imposed by deep potential wells or real and submerged barriers that appear in the lowest reaction channel under the present conditions. Note also that the present findings align with previous work, e.g., by Gingell \textit{et al.}~\cite{gingell10a} on reactions of halomethanes with Ca$^+$, that also observed increased reaction rates in excited states of Ca$^+$ close to those predicted by capture models.

Interestingly, the formation rate coefficients of C$_2$H$_2$CaCl$^+$ ($k_2$, P2) show no clear dependence on laser detuning, remaining approximately constant across different Ca$^+$ electronic state populations (see Table \ref{tbl:rate_constants}). 
Several factors could potentially explain this behaviour. First, the generally low product yields and experimental uncertainties associated with this channel might obscure underlying state-dependent trends. Second, in the excited states, our PES reveals that the P2 channel likely faces competition from multiple pathways including the barrierless CaCl$^+$ formation pathway (P1) and the energetically now open P1$'$ channel. Consequently, even as the excited-state population increases, these competing pathways may effectively limit C$_2$H$_2$CaCl$^+$ production.

The key difference between products P1 and P1$'$ lies in the structure of the neutral C$_2$H$_2$Cl radical formed alongside CaCl$^+$. Unlike the planar C--C--Cl linear structure observed in the cationic form, the neutral C$_2$H$_2$Cl radical can exist as three distinct isomers: 1,1-C$_2$H$_2$Cl (vinylidene structure) and the \textit{cis}/\textit{trans} conformers of 1,2-C$_2$H$_2$Cl. Previous theoretical~\cite{bai25a,li06a,zhang20a} and experimental studies~\cite{gao07a} investigating the reaction of Cl with C$_2$H$_2$ have demonstrated similar hydrogen migration processes, identifying these three C$_2$H$_2$Cl isomeric structures as either products or reaction intermediates. 
In Ca$^+$($^2$S$_{1/2}$) reactions, the neutral C$_2$H$_2$Cl radical is likely formed through dissociation from intermediate I3. The product distribution may favour \textit{cis}-C$_2$H$_2$Cl radicals due to geometric constraints during the dissociation process, although subsequent isomerisation to the more thermodynamically stable \textit{trans}-C$_2$H$_2$Cl form~\cite{bai25a,zhang20a} may also occur. In contrast, excited-state reactions may yield the 1,1-C$_2$H$_2$Cl isomer, which represents the global minimum on the potential energy surface.

\section{Conclusions}
In this study, we have investigated isomer- and state-dependent gas-phase reactions of Coulomb-crystallised Ca$^+$ ions with \textit{cis}- and \textit{trans}-1,2-dichloroethene under controlled conditions. The reactions exhibit clear isomer-specific behaviour, with \textit{cis}-DCE consistently showing higher reactivity than its \textit{trans} counterpart by approximately 20--30\%. This systematic difference is qualitatively predicted by capture theory and attributed to the role of the different electrostatic moments in governing the long-range ion--molecule interactions of the two isomers. The observation that the \textit{cis}/\textit{trans} reactivity ratio remains nearly constant across different Ca$^+$ electronic-state populations further demonstrates that these geometric effects persist regardless of the electronic configuration of the ion.

Our state-selective measurements have uncovered differences between ground and excited state reactions. While excited-state Ca$^+$ exhibits capture-limited behaviour with rate coefficients matching capture-theory predictions, ground-state reactions proceed at approximately half the predicted rate. This suppression of ground-state reactivity was tentatively attributed to the formation of a deeply bound reactive complex that can decay to either the reactants or the products with approximately equal probability.

The formation of C$_2$H$_2$CaCl$^+$ appears to exhibit approximately constant rate coefficients across different electronic state populations, possibly due to competition with other reaction pathways. Additionally, we observed the formation of secondary products, particularly C$_2$H$_2$CaCl$_3^+$ arising from sequential reactions. 

The present results underline the capability of Coulomb-crystallised ions as a platform for simultaneously studying state- and isomer-specific effects in ion--molecule reactions. Beyond providing quantitative benchmarks for theoretical models, our findings emphasise how long-range interactions interplay with short-range dynamical behaviour involving multiple reaction and energy-redistribution pathways. The insights gained here serve to deepen our understanding of how electronic state population and molecular geometry combine to influence reaction outcomes in ion--molecule systems.

\section*{Author contributions}

L.X. performed the experiments, analysed the data and conducted the quantum-chemical calculations. R.K. performed the OBE simulations of Ca$^+$ state populations. J.T. and S.W. supervised the project. L.X. drafted the manuscript, all authors contributed to the writing.

\section*{Conflicts of interest}
There are no conflicts to declare.

\section*{Data availability}

The primary data that support the findings of this study are available on Zenodo with the identifier doi:10.5281/zenodo.17084486.
%XXX.

\section*{Acknowledgements}

We acknowledge funding from the Swiss National Science Foundation, project grant nr. IZCOZ0\_189907 and Ambizione grant nr. PZ00P2 208818 (J.T.), as well as from the University of Basel.

%\bibliographystyle{tfo}
%\bibliography{DCE_Ca+_bibdesk}

\begin{thebibliography}{62}
\providecommand{\url}[1]{\texttt{#1}}
\providecommand{\urlprefix}{URL }

\bibitem{mcguire20a}
B.A. McGuire, O. Asvany, S. Brünken and S. Schlemmer,  Nat. Rev. Phys.  \textbf{2}, 402--410 (2020).

\bibitem{shuman15a}
N.S. Shuman, D.E. Hunton and A.A. Viggiano,  Chem. Rev.  \textbf{115}, 4542--4570 (2015).

\bibitem{williams02a}
K.L. Williams, I.T. Martin and E.R. Fisher,  J. Am. Soc. Mass Spectrom.  \textbf{13}, 518--529 (2002).

\bibitem{meyer17a}
J. Meyer and R. Wester,  Annu. Rev. Phys. Chem.  \textbf{68}, 333--353 (2017).

\bibitem{adams76a}
N.G. Adams and D. Smith,  Int. J. Mass Spectrom. Ion Phys.  \textbf{21}, 349--359 (1976).

\bibitem{armentrout02a}
P.B. Armentrout,  J. Am. Soc. Mass. Spectrom.  \textbf{13}, 419--434 (2002).

\bibitem{potapov17a}
A. Potapov, A. Canosa, E. Jiménez and B. Rowe,  Angew. Chem. Int. Ed.  \textbf{56}, 8618--8640 (2017).

\bibitem{allmendinger16a}
P. Allmendinger, J. Deiglmayr, K. H\"{o}veler, O. Schullian and F. Merkt,  J. Chem. Phys  \textbf{145}, 244316 (2016).

\bibitem{allmendinger16b}
P. Allmendinger, J. Deiglmayr, O. Schullian, K. H{\"o}veler, J.A. Agner, H. Schmutz and F. Merkt,  ChemPhysChem  \textbf{17} (22), 3596--3608 (2016).

\bibitem{martins25a}
F.B.V. Martins, H. Schmutz, J.A. Agner, V. Zhelyazkova and F. Merkt,  Phys. Rev. Lett.  \textbf{134}, 123401 (2025).

\bibitem{willitsch12a}
S. Willitsch,  Int. Rev. Phys. Chem.  \textbf{31}, 175 (2012).

\bibitem{willitsch17a}
S. Willitsch,  Adv. Chem. Phys.  \textbf{162}, 307 (2017).

\bibitem{roth06b}
B. Roth, P. Blythe, H. Wenz, H. Daerr and S. Schiller,  Phys. Rev. A  \textbf{73}, 042712 (2006).

\bibitem{yang18a}
T. Yang, A. Li, G.K. Chen, C. Xie, A.G. Suits, W.C. Campbell, H. Guo and E.R. Hudson,  J. Phys. Chem. Lett.  \textbf{9}, 3555--3560 (2018).

\bibitem{molhave00a}
K. M{\o}lhave and M. Drewsen,  {Phys. Rev. A}  \textbf{62}, 011401 (2000).

\bibitem{staanum08a}
P.F. Staanum, K. H{\o}jbjerre, R. Wester and M. Drewsen,  Phys. Rev. Lett.  \textbf{100}, 243003 (2008).

\bibitem{willitsch08a}
S. Willitsch, M.T. Bell, A.D. Gingell, S.R. Procter and T.P. Softley,  Phys. Rev. Lett.  \textbf{100}, 043203 (2008).

\bibitem{gingell10a}
A.D. Gingell, M.T. Bell, J.M. Oldham, T.P. Softley and J.N. Harvey,  J. Chem. Phys.  \textbf{133}, 194302 (2010).

\bibitem{chang13a}
Y.P. Chang, K. Dlugolecki, J. K{\"u}pper, D. R{\"o}sch, D. Wild and S. Willitsch,  Science  \textbf{342}, 98 (2013).

\bibitem{schmid19a}
P.C. Schmid, M.I. Miller, J. Greenberg, T.L. Nguyen, J.F. Stanton and H.J. Lewandowski,  Mol. Phys.  \textbf{117}, 3036--3042 (2019).

\bibitem{kilaj23a}
A. Kilaj, S. Käser, J. Wang, P. Stra\v{n}ák, M. Schwilk, L. Xu, O.A. von Lilienfeld, J. Küpper, M. Meuwly and S. Willitsch,  Phys. Chem. Chem. Phys.  \textbf{25}, 13933--13945 (2023).

\bibitem{roth08a}
B. Roth, D. Offenberg, C.B. Zhang and S. Schiller,  Phys. Rev. A  \textbf{78}, 042709 (2008).

\bibitem{hall13b}
F.H.J. Hall, M. Aymar, M. Raoult, O. Dulieu and S. Willitsch,  Mol. Phys.  \textbf{111}, 1683--1690 (2013).

\bibitem{mahdian21a}
A. Mahdian, A. Kr{\"u}kow and J.H. Denschlag,  New J. Phys.  \textbf{23}, 065008 (2021).

\bibitem{mohammadi21a}
A. Mohammadi, A. Kr{\"u}kow, A. Mahdian, M. Dei{\ss}, J. P{\'e}rez-R{\'\i}os, H. da~Silva~Jr, M. Raoult, O. Dulieu and J. Hecker~Denschlag,  Phys. Rev. Res.  \textbf{3}, 013196 (2021).

\bibitem{zhelyazkova20a}
V. Zhelyazkova, F.B.V. Martins, J.A. Agner, H. Schmutz and F. Merkt,  Phys. Rev. Lett.  \textbf{125}, 263401 (2020).

\bibitem{tsikritea22a}
A. Tsikritea, J.A. Diprose, T.P. Softley and B.R. Heazlewood,  J. Chem. Phys.  \textbf{157}, 060901 (2022).

\bibitem{clary84a}
D.C. Clary,  Mol. Phys.  \textbf{53}, 3--21 (1984).

\bibitem{doerfler19a}
A.D. D{\"o}rfler, P. Eberle, D. Koner, M. Tomza, M. Meuwly and S. Willitsch,  Nat. Commun.  \textbf{10}, 5429 (2019).

\bibitem{Kocheril2025a}
G.S. Kocheril, C. Zagorec-Marks and H.J. Lewandowski,  Phys. Rev. Lett.  \textbf{134}, 203401 (2025).

\bibitem{schmid20a}
P.C. Schmid, J. Greenberg, T.L. Nguyen, J.H. Thorpe, K.J. Catani, O.A. Krohn, M.I. Miller, J.F. Stanton and H.J. Lewandowski,  Phys. Chem. Chem. Phys.  \textbf{22}, 20303--20310 (2020).

\bibitem{kilaj21a}
A. Kilaj, J. Wang, P. Stra\v{n}ák, M. Schwilk, U. Rivero, L. Xu, O.A. von Lilienfeld, J. Küpper and S. Willitsch,  Nat. Commun.  \textbf{12}, 6047 (2021).

\bibitem{ploenes24a}
L. Ploenes, P. Straňák, A. Mishra, X. Liu, J. Pérez-Ríos and S. Willitsch,  Nat. Chem.  \textbf{16}, 1876--1881 (2024).

\bibitem{richardson24a}
V. Richardson, L. Alcock, N. Solem, D. Sundelin, C. Romanzin, R. Thissen, W.D. Geppert, C. Alcaraz, M. Polášek, B.R. Heazlewood, Q. Autret and D. Ascenzi,  J. Phys. Chem. Lett.  \textbf{15}, 10888--10895 (2024).

\bibitem{zagorecmarks24a}
C. Zagorec-Marks, G.S. Kocheril, O.A. Krohn, T. Kieft, A. Karpinska, T.P. Softley and H.J. Lewandowski,  Faraday Discuss.  \textbf{251}, 125--139 (2024).

\bibitem{haynes16a}
W.M. Haynes, editor, \emph{CRC Handbook of Chemistry and Physics}, 97th ed.   (CRC Press, Boca Raton, FL, 2016).

\bibitem{zhi21a}
Y. Zhi, J. Hu, J.C. Xie and S.X. Tian,  J. Phys. Chem. A  \textbf{125}, 2573--2580 (2021).

\bibitem{mikhailov06a}
V.A. Mikhailov, M.A. Parkes, R.P. Tuckett and C.A. Mayhew,  J. Phys. Chem. A  \textbf{110}, 5760--5771 (2006).

\bibitem{rebrion88b}
C. Rebrion, J.B. Marquette, B.R. Rowe, C. Chakravarty, D.C. Clary, N.G. Adams and D. Smith,  J. Phys. Chem.  \textbf{92}, 6572--6574 (1988).

\bibitem{roesch16a}
D. R\"osch, H. Gao, A. Kilaj and S. Willitsch,  EPJ Tech. Instrum.  \textbf{3}, 5 (2016).

\bibitem{kilaj18a}
A. Kilaj, H. Gao, D. R{\"o}sch, U. Rivero, J. K{\"u}pper and S. Willitsch,  Nat. Commun.  \textbf{9} (2018).

\bibitem{xu24a}
L. Xu, J. Toscano and S. Willitsch,  Phys. Rev. Lett.  \textbf{132}, 083001 (2024).

\bibitem{bartmess83a}
J.E. Bartmess and R.M. Georgiadis,  Vacuum  \textbf{33}, 149--153 (1983).

\bibitem{nakao75a}
F. Nakao,  Vacuum  \textbf{25}, 431--435 (1975).

\bibitem{jeffers72a}
P.M. Jeffers,  J. Phys. Chem.  \textbf{76}, 2829--2832 (1972).

\bibitem{gioumousis58a}
G. Gioumousis and D.P. Stevenson,  J. Chem. Phys.  \textbf{29}, 294 (1958).

\bibitem{su73b}
T. Su and M.T. Bowers,  Int. J. Mass Spectrom. Ion Phys.  \textbf{12}, 347--356 (1973).

\bibitem{adamo99a}
C. Adamo and V. Barone,  J. Chem. Phys.  \textbf{110}, 6158--6170 (1999).

\bibitem{grimme11a}
S. Grimme, S. Ehrlich and L. Goerigk,  J. Comput. Chem.  \textbf{32}, 1456--1465 (2011).

\bibitem{becke05a}
A.D. Becke and E.R. Johnson,  J. Chem. Phys.  \textbf{123}, 154101 (2005).

\bibitem{raghavachari89a}
K. Raghavachari, G.W. Trucks, J.A. Pople and M. Head-Gordon,  Chem. Phys. Lett.  \textbf{157}, 479--483 (1989).

\bibitem{dunning89a}
J. Thom H.~Dunning,  J. Chem. Phys.  \textbf{90}, 1007 (1989).

\bibitem{g16}
M.J. Frisch, G.W. Trucks, H.B. Schlegel, G.E. Scuseria, M.A. Robb, J.R. Cheeseman, G. Scalmani, V. Barone, G.A. Petersson, H. Nakatsuji, X. Li, M. Caricato, A.V. Marenich, J. Bloino, B.G. Janesko, R. Gomperts, B. Mennucci, H.P. Hratchian, J.V. Ortiz, A.F. Izmaylov, J.L. Sonnenberg, D. Williams-Young, F. Ding, F. Lipparini, F. Egidi, J. Goings, B. Peng, A. Petrone, T. Henderson, D. Ranasinghe, V.G. Zakrzewski, J. Gao, N. Rega, G. Zheng, W. Liang, M. Hada, M. Ehara, K. Toyota, R. Fukuda, J. Hasegawa, M. Ishida, T. Nakajima, Y. Honda, O. Kitao, H. Nakai, T. Vreven, K. Throssell, J.A. Montgomery, {Jr.}, J.E. Peralta, F. Ogliaro, M.J. Bearpark, J.J. Heyd, E.N. Brothers, K.N. Kudin, V.N. Staroverov, T.A. Keith, R. Kobayashi, J. Normand, K. Raghavachari, A.P. Rendell, J.C. Burant, S.S. Iyengar, J. Tomasi, M. Cossi, J.M. Millam, M. Klene, C. Adamo, R. Cammi, J.W. Ochterski, R.L. Martin, K. Morokuma, O. Farkas, J.B. Foresman and D.J. Fox, \emph{Gaussian 16 {R}evision {C}.01}, Gaussian Inc., Wallingford CT, 2016.

\bibitem{marti03a}
R. Martí, in \emph{Handbook of Metaheuristics}, edited by Fred Glover and Gary~A. Kochenberger  (Springer, Boston, MA, 2003), pp. 355--368.

\bibitem{chen23a}
Z.X. Chen, J. Hu, Y. Zhi, C.X. Wu and S.X. Tian,  Chin. J. Chem. Phys.  \textbf{36}, 509--516 (2023).

\bibitem{okada02a}
K. Okada, M. Wada, L. Boesten, T. Nakamura, I. Katayama and S. Ohtani,  J. Phys. B  \textbf{36}, 33--46 (2003).

\bibitem{wu24a}
Z. Wu, S. Walser, V. Podlesnic, M. Isaza-Monsalve, E. Mattivi, G. Mu, R. Nardi, P. Gniewek, M. Tomza, B.J. Furey and P. Schindler,  J. Chem. Phys.  \textbf{161}, 044304 (2024).

\bibitem{bodi11a}
A. Bodi, W.R. Stevens and T. Baer,  J. Phys. Chem. A  \textbf{115}, 726--734 (2011).

\bibitem{bai25a}
Y. Bai, Y.L. Fu, J. Qi, L. Liu, X. Lu, Y.C. Han, D.H. Zhang and B. Fu,  Nat. Commun.  \textbf{16}, 2732 (2025).

\bibitem{li06a}
J.L. Li, C.Y. Geng, X.R. Huang, J.H. Zhan and C.C. Sun,  Chem. Phys.  \textbf{331}, 42--54 (2006).

\bibitem{zhang20a}
L. Zhang, D.G. Truhlar and S. Sun,  Proc. Natl. Acad. Sci.  \textbf{117}, 5610--5616 (2020).

\bibitem{gao07a}
Y. Gao, I.M. Alecu, P.C. Hsieh, A. McLeod, C. McLeod, M. Jones and P. Marshall,  Proc. Combust. Inst.  \textbf{31}, 193--200 (2007).

\bibitem{craig71a}
N.C. Craig, L.G. Piper and V.L. Wheeler, J. Phys. Chem. \textbf{75}, 1453 (1971).

\bibitem{kanakaraju02a}
R. Kanakaraju, K. Senthilkumar and P. Kolandaivel, J. Mol. Struct. (Theochem) \textbf{589-590}, 95 (2002).

\end{thebibliography}

\end{document}